\documentclass[         %
aps,                    
prd,                    
showpacs,               
nofootinbib,            
showkeys,               %
preprintnumbers,        %
floatfix]               
{revtex4}               
\usepackage{graphicx,longtable}

\begin{document}
\title{Impact of the Neutrino Magnetic Moment on 
the Neutrino Fluxes and the Electron Fraction in core-collapse
Supernovae}

\author{        A.~B. Balantekin}
\email{         baha@physics.wisc.edu}
\affiliation{  Department of Physics, University of Wisconsin\\
               Madison, Wisconsin 53706 USA }
\author{        C. Volpe}
\email{volpe@ipno.in2p3.fr}
\author{J. Welzel}
\email{   welzel@ipno.in2p3.fr}
\affiliation{Institut de Physique Nucl\'eaire, F-91406 Orsay cedex,
France}

\date{\today}
\begin{abstract}
We explore the effect of the neutrino magnetic moment on neutrino scattering
with matter in a core-collapse Supernova. We study the impact both on the neutrino fluxes and on the electron fraction. We find that sizeable modifications require
very large magnetic moments both for Dirac and Majorana neutrinos.
\end{abstract}
\medskip
\preprint{}
\maketitle
                                                      
\section{Introduction}

Core collapse supernovae are powerful neutrino sources of all flavours, since
their explosion produces a very intense burst of neutrinos 
with an energy of tens of MeV
on a timescale of several seconds. These expectations have been 
roughly confirmed 
by the explosion of the Supernova 1987A in the Large Magellanic Cloud.
Besides, their observation has helped to better constrain various 
neutrino properties
that still remain unknown. The measurement of a future galactic supernova
explosion or of the diffuse supernova neutrino background would be of great
interest both for our knowledge of neutrino properties and for unraveling the
supernova explosion mechanism since in present simulations the shock wave
fails to eject the mantle. While this keeps being an open issue, convection or
extra deposit of energy might be the key to solve
this question. Because 99 $\%$ of the energy is emitted by neutrinos, it is
likely that they might drive the shock wave out the star by some still unknown
mechanism.

Neutrinos also play a role in the rapid neutron
capture (r-process) nucleosynthesis scenario \cite{Burbidge:1957vc},
where
a good fraction of the heavier nuclei were formed. 
Although an astrophysical site of the r-process is not yet identified, 
one expects such sites to be associated with explosive phenomena since 
a large number of interactions are required to take place during a 
rather short time interval. Time-scale arguments based on meteoritic 
data imply that r-process nuclei may come from diverse sources 
\cite{Qian:1997kz}. Neutrino-driven wind models of neutron-rich material 
ejection following core-collapse supernovae indicate a possible site 
\cite{Woosley:1992ek,Woosley:1994ux,Takahashi:1994yz}. A key quantity 
for determining the r-process yields is the neutron-to-seed nucleus ratio, 
which in turn, is determined by the neutron-to-proton ratio at freeze-out. 
The neutrino-induced process $\nu_e + n \rightarrow p + e^-$, operating 
during or immediately after the freeze-out, could significantly alter 
neutron-to-proton ratio. During the epoch of alpha-particle formation almost 
all the protons and an equal amount of neutrons combine into alpha particles 
which have a large binding energy. This ``alpha effect'' reduces the number of 
free neutrons taking place in the r-process, pushing the electron fraction 
close to $Y_e = 0.5$ \cite{Fuller:1995ih,Meyer:1998sn}. 
One possible scheme 
to reduce the impact of the alpha effect is to reduce the electron neutrino 
flux. This should happen at relatively far away from the vicinity 
of the neutron star so that neutrino heating already can have taken place. 
Oscillations between active flavors will only increase the conversion of 
neutrons to protons since mu and tau neutrinos are likely to have higher 
energies than electron neutrinos to begin with (The exact hierarchy of 
neutrino energies depends on the details of microphysics 
\cite{Hannestad:1997gc,Buras:2002wt}). 
However, a reduction of the electron neutrino flux can be achieved by 
oscillation of electron neutrinos into sterile neutrinos 
\cite{McLaughlin:1999pd,Caldwell:1999zk,Fetter:2002xx}. 

In this article we study the impact of the neutrino magnetic moment on both
the neutrino fluxes and the electron fraction in a core-collapse supernova
environment. The cases of Dirac and Majorana neutrinos are analyzed, inducing
active-sterile and active-active conversions respectively. 
In our 
considerations we ignore the possibility of the presence of large magnetic 
fields near the supernova core. Such magnetic fields could cause additional 
transformations between neutrino flavors via spin-flavor precession scenarios 
\cite{Nunokawa:1996gp,Ando:2003pj}.
We also ignore neutrino-neutrino interactions 
\cite{Duan:2006jv,Balantekin:2006tg,Hannestad:2006nj}.
Note that the impact of the neutrino magnetic moment in astrophysical and
cosmological contexts has been discussed in various works, e.g.
\cite{Raffelt:1990pj} for red-giant cooling, in \cite{Morgan:1981zy} 
for big-bang nucleosynthesis, 
for solar neutrinos \cite{Balantekin:2004tk} and
in \cite{Barbieri:1988nh,Lattimer:1988mf,Notzold:1988kz} 
for core-collapse Supernovae.

The plan of this paper is as follows. In the next section we summarize 
properties of the post core-bounce supernova in the nucleosynthesis epoch, 
discuss neutrino magnetic moment scattering in the pertinent plasma. In 
sections 3 and 4 we analyze the case of Dirac and Majorana neutrinos, 
respectively. 
We summarize our results in Section 5 where we present our conclusions. 

\section{Neutrino elastic scattering via magnetic moment interaction in a 
core-collapse supernova} 

A heuristic description of the conditions of the neutrino-driven wind
in  post-core bounce supernova environment is outlined in Refs.
\cite{McLaughlin:1999pd} and \cite{Balantekin:2004ug}. Neutrino
magnetic  moment effects could be present inside\footnote{In fact a
sufficiently large  neutrino magnetic moment can cause significant
energy losses during the  core-collapse and neutron-star formation
epochs. Observational limits on  the reduction of the trapped lepton
number were also used the constrain  the neutrino magnetic moments
\cite{Barbieri:1988nh,Lattimer:1988mf,Notzold:1988kz,Ayala:1998qz}.
Our analysis and limits we obtain deal with later epochs when
nucleosynthesis may take place.}  and just above the proto-neutron
star. The medium  immediately above the neutron star is a degenerate
and  relativistic plasma (we have $T_{Fermi}\gg T \approx 10^{10}$K).
The effective photon mass is then~\cite{Braaten:1993jw}
\begin{equation}
\label{1}
m_{\gamma}^2(N_e,\,T) =
\frac{2\alpha}{\pi}\left(\mu^2+\frac{1}{3}\pi^2T^2\right)
\end{equation}
with $\mu$ the electronic chemical potential :
\begin{equation}
\label{2}
\mu =\left(\sqrt{\frac{p_F^6}{4}+\frac{\pi^6T^6}{27}} +
\frac{p_F^3}{2}\right)^{1/3}-\left(\sqrt{\frac{p_F^6}{4} +
\frac{\pi^6T^6}{27}}-\frac{p_F^3}{2}\right)^{1/3},
\end{equation}
where the Fermi momentum is given by
\begin{equation}
\label{3}
p_F^3 = 3\pi^2N_e(r)
\end{equation}
In these equations the electron number density, $N_e$ is related to
the  matter density, $\rho$, as
\begin{equation}
\label{4}
N_e(r) = Y_e(r) \times \rho(r) / m_N,
\end{equation}
where $Y_e$ is the electron fraction and $m_N$ is the nucleon mass.
At the surface of the proto-neutron star, the density profile falls off
steeply over few kms.  For regions sufficiently removed from  the
proto-neutron star, density goes over to the  neutrino-driven wind
solution ($\sim 1/r^3$). In our calculations we  adopted the density
profile of Ref.\cite{McLaughlin:1999pd}  (with entropy $S=70 k_B$).

\begin{figure}
\centerline{\includegraphics[height=8.5cm,angle=-90]{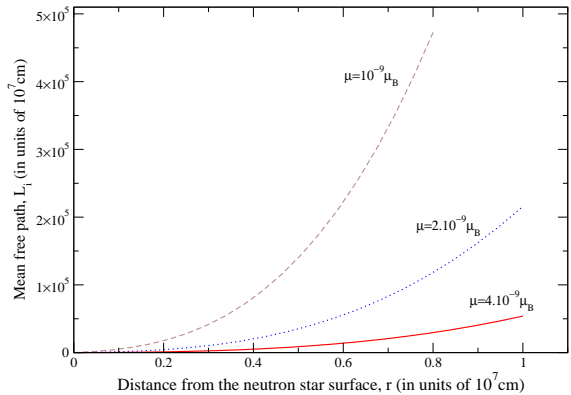}}
\caption{Neutrino mean free path $L_e$ as a function of the distance from
the neutron star $r$, both in units of $10^7$ cm.  From top to bottom,
$\mu_{\nu}=1\times10^{-9}\mu_B,\ 2\times10^{-9}\mu_B,\
4 \times 10^{-9}\mu_B$ respectively.\label{fig:2}}
\end{figure}

Magnetic contribution to the differential cross section for elastic
neutrino  scattering on electron is
\begin{equation}
\label{5}
\frac{d\sigma}{dt} = \left( \sum_f \mu_{if}^2 \right)  \frac{\pi
\alpha^2}{m_e^2} \frac{s + t - m_e^2}{(t-m_{\gamma}^2)(s-m_e^2)}.
\end{equation}
We ignore the contributions from the weak neutral-current scattering
which  preserves both the neutrino flavor and chirality.  Note that,
in the most general case, we sum over the contributions coming  from
both diagonal and transition magnetic moments since   the magnetic
scattering can produce any neutrino flavor. Hence for an  electron
neutrino in the initial state one has $\sum_f \mu_{if}^2 =  \mu_{ee}^2
+ \mu_{e \mu}^2 + \mu_{e \tau}^2$.  (For Majorana  neutrinos there are
no diagonal $\mu_{ii}$ magnetic moments).  Integrating Eq. (\ref{5})
we obtain the total cross section~\footnote{We note that the following
series expansion is useful for understanding  the convergence behavior
of this cross section:
\begin{equation}
\label{7}
\sigma = \left( \sum_f \mu_{if}^2 \right) \left( \frac{\pi
\alpha^2}{m_e^2} \right) \sum_{n=1}^{\infty} \frac{x^n}{n+1},
\end{equation}
where
\begin{equation}
\label{8}
x =  \frac{2m_eE_{\nu}}{2m_eE_{\nu}+m_{\gamma}^2}.
\end{equation}}
\begin{equation}
\label{6}
\sigma = \left( \sum_f \mu_{if}^2 \right) \frac{\pi \alpha^2}{m_e^2}
\left[ \left( 1 + \frac{m_{\gamma}^2}{2m_eE_{\nu}} \right) \times \log
\left( \frac{2m_eE_{\nu}+m_{\gamma}^2}{m_{\gamma}^2} \right) -
1\right].
\end{equation}
Since the effective photon mass can be large in our case, we keep the
constant term in Eq. (\ref{6}), usually ignored in the  literature.
The neutrino mean free path, $L_i$ is then
\begin{equation}
\label{9}
L_i =\frac{1}{\sigma(r,\,E_{\nu},\,\sum_f\mu_{if}^2) N_e(r)} .
\end{equation}
In Figure \ref{fig:2} we display the behavior of the neutrino mean free path as
a  function of the distance $r$ from the neutron star surface, for
various magnetic moment values. It can be seen that $L_e$ is very large, 
and therefore the magnetic moment interactions will be significant only 
very close to the proto-neutron star surface. 

The presence of the neutrino magnetic moment modifies the fluxes:
\begin{equation}\label{e:13bis}
\tilde{\phi}(E_{\nu},r)= 
\phi(E_{\nu}) {\cal N}_{\nu_e,\bar{\nu}_e}(E_{\nu},r)
\end{equation}
with $\phi(E_{\nu})$  the neutrino fluxes at the neutrinosphere
that we take as Fermi-Dirac distributions.\footnote{Note that different
neutrino flux shapes (power law) have been pointed out recently
\cite{Keil:2002in}. }
The electron
(anti)neutrino fraction ${\cal N}_{\nu_e,\bar{\nu}_e}$ is determined
by solving the evolution of the neutrino amplitudes or probabilities
in matter including the extra terms due to  the magnetic
interaction. We give these equations in the next sections both for
Dirac and Majorana neutrinos.
As far as the equilibrium  electron fraction in the supernova is concerned,
it is given by
\begin{equation}
\label{e:12}
Y_e(r) \sim \frac{1}{1+\frac{\lambda_{\bar{\nu}_e p}(r)}{\lambda_{\nu_e
n}(r)}}.
\end{equation} 
in the absence of a significant number of alpha particles, as the 
magnetic moment acts at early times, very close to the neutron star 
surface (Figure \ref{fig:2}). 
In Eq. (\ref{e:12}) $\lambda_{\bar{\nu}_e p}$ is the rate of the
reaction  $\bar{\nu}_e + p \rightarrow n +e^+$ producing neutrons and
$\lambda_{\nu_e n}$ is the rate of the reaction  $\nu_e + n
\rightarrow p + e^-$ destroying neutrons.  These rates are given by
\begin{equation}\label{e:13}
\lambda_{\nu_en,\bar{\nu}_ep}(r)= \int \sigma_{weak}(E_{\nu})
\phi(E_{\nu},r) {\cal N}_{\nu_e,\bar{\nu}_e}(E_{\nu},r) dE_{\nu}
\end{equation}
\noindent 
where the cross section is $\sigma_{weak}(E_{\nu})= 9.6 \times 10^{-44}
(E_{\nu} \pm 1.293)^2$ cm$^{2}$  for neutrinos (minus for
anti-neutrinos).

\section{Dirac neutrinos}
Let us first discuss Dirac neutrinos in the case of two flavors to
illustrate the salient features of the evolution. The evolution
equation of the neutrino amplitudes, including both the standard
matter (MSW) effect and the magnetic moment interaction, is given by
\begin{equation}
i\frac{\partial}{\partial r} \left[\begin{array}{cc}
\Psi_{\nu_e}(E_{\nu},r) \\ \\ \Psi_{\nu_{\mu}}(E_{\nu},r)
\end{array}\right] = \left[\begin{array}{cc} \varphi(r)-\frac{i}{2L_e}
& \frac{\delta m^2}{4 E_{\nu}}\sin{2\theta_v} \\ \\  \frac{\delta
m^2}{4 E_{\nu}}\sin{2\theta_v} & -\varphi(r)-\frac{i}{2L_{\mu}}
\end{array}\right]
\left[\begin{array}{cc} \Psi_{\nu_e}(E_{\nu},r) \\ \\
\Psi_{\nu_{\mu}}(E_{\nu},r)
  \end{array}\right]\,,
\label{10}
\end{equation}
with $\theta_v$ being the neutrino vacuum mixing angle, $\delta m^2$
the square mass difference, $L_i$ from Eq.(\ref{9}) and where
\begin{equation}
  \label{11}
  \varphi(r) = \frac{1}{4 E_{\nu}}\left( 2 \sqrt{2}\ G_F N_e(r)
E_{\nu} -  \delta m^2 \cos{2\theta_v} \right) .
\end{equation}
In Eq.(\ref{10}) the term due to the neutrino magnetic moment is such
that the electron survival probability is suppressed by a $1/e$ factor
at a distance in the star equal to one mean free path. Indeed, since
the magnetic scattering produces wrong-chirality  (sterile) states,
such an equation produces a net loss of flux from all the channels.

Rewriting Eq.(\ref{10}) for three flavors
we have 
\begin{equation}
i\frac{\partial}{\partial r} \left[\begin{array}{cc}
\Psi_{\nu_e}(E_{\nu},r) \\\Psi_{\nu_{\mu}}(E_{\nu},r)
\\\Psi_{\nu_{\tau}}(E_{\nu},r)\end{array}\right] =
\left( \hat{H}_{\rm MSW} +
\left[\begin{array}{ccc} - \frac{i}{2L_{e}}&0&0\\ 0& -\frac{i}{2L_{\mu}}& 0\\
0&0&-\frac{i}{2L_{\tau}}
\end{array}\right] 
\right)
\left[\begin{array}{cc} \Psi_{\nu_e}(E_{\nu},r)
\\\Psi_{\nu_{\mu}}(E_{\nu},r)
\\\Psi_{\nu_{\tau}}(E_{\nu},r)\end{array}\right] .
\label{e:14}
\end{equation}
The region where the
density is large enough to render the interaction due to the magnetic
moment effective is the high density region very close to the neutron
star surface, far from the MSW resonances~\footnote{For this reason, 
the neutrino mass hierarchy will not influence our results.}.  In 
this region the term $H_{\rm MSW}$ in Eq. (\ref{e:14}) contributes 
very little, which we verified numerically.

\begin{figure}[!]
\centerline{\includegraphics[height=8.5cm,angle=-90]{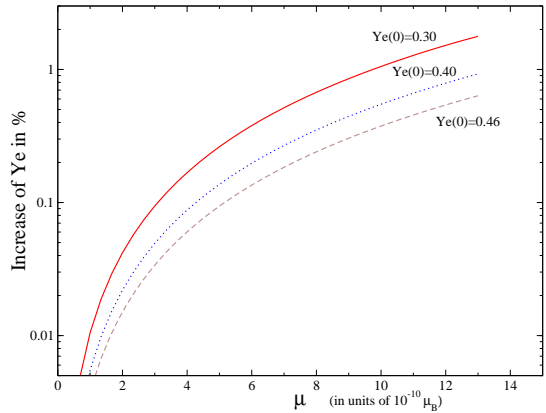}}
\caption{{\sc Case of Dirac neutrinos :} Increase of the electron
fraction in percentage, $(Y_e(r)-Y_e(r=0))/Y_e(r=0)$, evaluated at a
distance of $r=4$ km from the neutron star surface, as a function of
the neutrino magnetic moment $\mu=(\sum_f\mu_{e f}^2)^{1/2}$. }
\label{figYeDirac}
\end{figure}

We have solved equation~(\ref{e:14}), calculated 
the neutrino fluxes Eq.(\ref{e:13bis})  and the reaction rates
Eqs.(\ref{e:12}-\ref{e:13}) both for neutrinos and 
antineutrinos~\footnote{Anti-neutrinos obey the same type of equations 
but with $H_{\rm MSW}$ modified by a minus sign in front of the MSW 
potential.}. We find that the effect on the neutrino fluxes
is not significant even for
very large values of the neutrino magnetic moment. 
Figure~\ref{figYeDirac} presents the variation of the electron
fraction, i.e. $(Y_e(r)-Y_e(r=0))/Y_e(r=0)$, in percentage at a
distance of $r=4$ km from the neutron star surface, where the magnetic
moment interaction become ineffective. The results are given 
for different hierarchies of neutrinos temperatures~\footnote{Note 
that different sets of neutrinos temperatures can lead to the same value 
of $Y_e(0)$, {\it cf.} Eq~\ref{e:12}-\ref{e:13}. However, for a given 
$Y_e(0)$, we checked that our results depend only a little on the 
corresponding hierarchies. Moreover, the dependence of our results on 
$\nu_{\mu},\nu_{\tau}$ temperatures is not significant. They have 
been fixed at the value $T_{\nu_x,\bar{\nu}_x}=7.5$ MeV.}, {\it i.e.} 
different electron fraction 
$Y_e(r=0)$, given in table~\ref{tableTnu}. One can see that the electron
fraction increases as the magnetic moment $\mu^2=\sum_f\mu_{ef}^2$
gets larger.  For Dirac neutrinos it is clear that the  presence of
the neutrino magnetic moment converts both electron neutrino and
electron antineutrino fluxes into sterile states. Hence  it lowers
both of these rates; but, since the electron anti-neutrinos are more
energetic, it lowers the neutron production rate more because of the
cross section behavior Eq.(\ref{6}). Then, the ratio 
$\lambda_{\bar{\nu}_e p}(r)/\lambda_{\nu_e
n}(r)$ decreases and $Y_e(r)$ increases.  A magnetic moment as 
large as $\mu = 10^{-9}~\mu_B$ (Bohr magnetons) induces an increase 
of $Y_e$ of $1 \%$. 
However the variation $Y_e$ strongly depends on the $\mu_{e
f}$. For example a value of $3 \times 10^{-10}~\mu_B$ produces an
increase up to at most $0.1 \%$, while for the present experimental 
upper-limit of $(\sum_f\mu_{e f}^2)^{1/2} \leq 0.74\times 
10^{-10}~\mu_B$~\cite{Yao:2006px} one gets a small effect of less 
than 0.01$\%$~\footnote{One should also 
note that such large values of magnetic moment would cause the neutron 
star lose its energy too fast to begin with.}. One also see that, for a $Y_e(r=0)$ closer to the 
critical value 0.5, the effect becomes smaller. 

\begin{center}
\begin{table}[h!]
\begin{tabular}{c|c|c|c|c}
Electron fraction $Y_e(0)$ & $T_{\nu_e}$ (MeV)& 
$T_{\bar{\nu}_e}$ (MeV) & $\left<E_{\nu_e}\right>$ (MeV) & 
$\left<E_{\bar{\nu}_e}\right>$ (MeV)\\ \hline
0.30 & 2.1 & 7.1 & 6.6 & 22.4 \\ \hline
0.40 & 3.0 & 6.0 & 9.4 & 18.9 \\ \hline
0.46 & 3.5 & 5.7 & 11 & 18 \\ \hline
\end{tabular}
\caption{Electron fraction at the proto-neutron star surface and 
one possible set of associated neutrino temperature hierarchies.}
\label{tableTnu}
\end{table}
\end{center}

\section{Majorana Neutrinos}
In the Majorana case, the evolution equations including the neutrino
magnetic moment effects  involve both neutrinos and anti-neutrinos
(the $\nu_R$'s are no more sterile).  Since in the Dirac case we
have checked that standard matter effects modify little the neutrino
evolution at the region where the magnetic field effects are important, 
it is sufficient to consider the evolution equation for all neutrino 
number fraction ${\cal
N}_{\nu_{x}}$ including the
neutrino magnetic moment effect only :
\begin{equation}\label{e:matevo}
{\partial \over \partial r} \left[\begin{array}{cccccc} {\cal
N}_{\nu_{e L}} \\  {\cal N}_{\nu_{\mu L}} \\ {\cal N}_{\nu_{\tau L}}
\\  {\cal N}_{\nu_{e R}} \\  {\cal N}_{\nu_{\mu R}} \\ {\cal
N}_{\nu_{\tau R}}  \end{array}\right] =
\left[\begin{array}{cccccc}\label{e:15} -\lambda_1-\lambda_2 & 0 & 0 &
0 & \lambda_1 & \lambda_2 \\ 0 & -\lambda_1-\lambda_3 & 0 & \lambda_1
&0 & \lambda_3   \\ 0 & 0 & -\lambda_2-\lambda_3& \lambda_2 &
\lambda_3 & 0 \\ 0& \lambda_1 & \lambda_2 & -\lambda_1-\lambda_2 & 0 &
0\\ \lambda_1 &0 & \lambda_3& 0 & -\lambda_1-\lambda_3& 0 \\
\lambda_2& \lambda_3 & 0& 0 & 0& -\lambda_2-\lambda_3 \\
\end{array}\right]
\left[\begin{array}{cccccc}  {\cal N}_{\nu_{e L}} \\  {\cal
N}_{\nu_{\mu L}} \\ {\cal N}_{\nu_{\tau L}} \\  {\cal N}_{\nu_{e R}}
\\  {\cal N}_{\nu_{\mu R}} \\ {\cal N}_{\nu_{\tau R}}
\end{array}\right]
\end{equation}
with $\lambda_1=1/L_{e\mu}$, $\lambda_2=1/L_{e\tau}$,
$\lambda_3=1/L_{\mu\tau}$. The mean free path $L_{if}$ include the
effect of the transition magnetic moments $\mu_{if}$. For each
species, there are four conversions : two contribute positively
(gain), two negatively (loss). For example, the left-handed electron
neutrino number gains from $\nu_{\mu R},\nu_{\tau R}\to\nu_{e L}$ and
looses from $\nu_{e L}\to\nu_{\mu R},\nu_{\tau R}$.

\begin{figure}[!]
\centerline{\includegraphics[height=8cm,angle=-90]{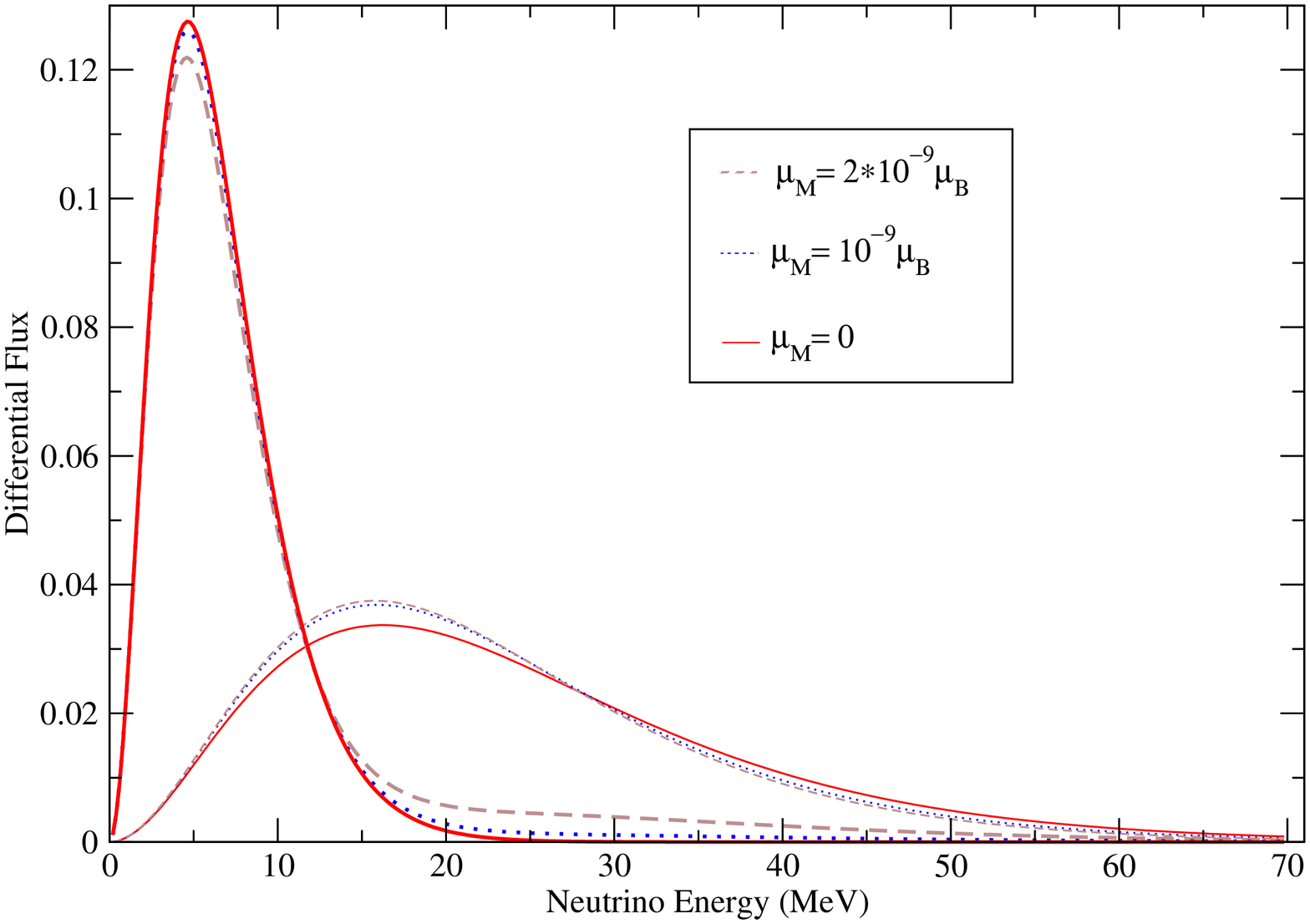}\includegraphics[height=8cm,angle=-90]{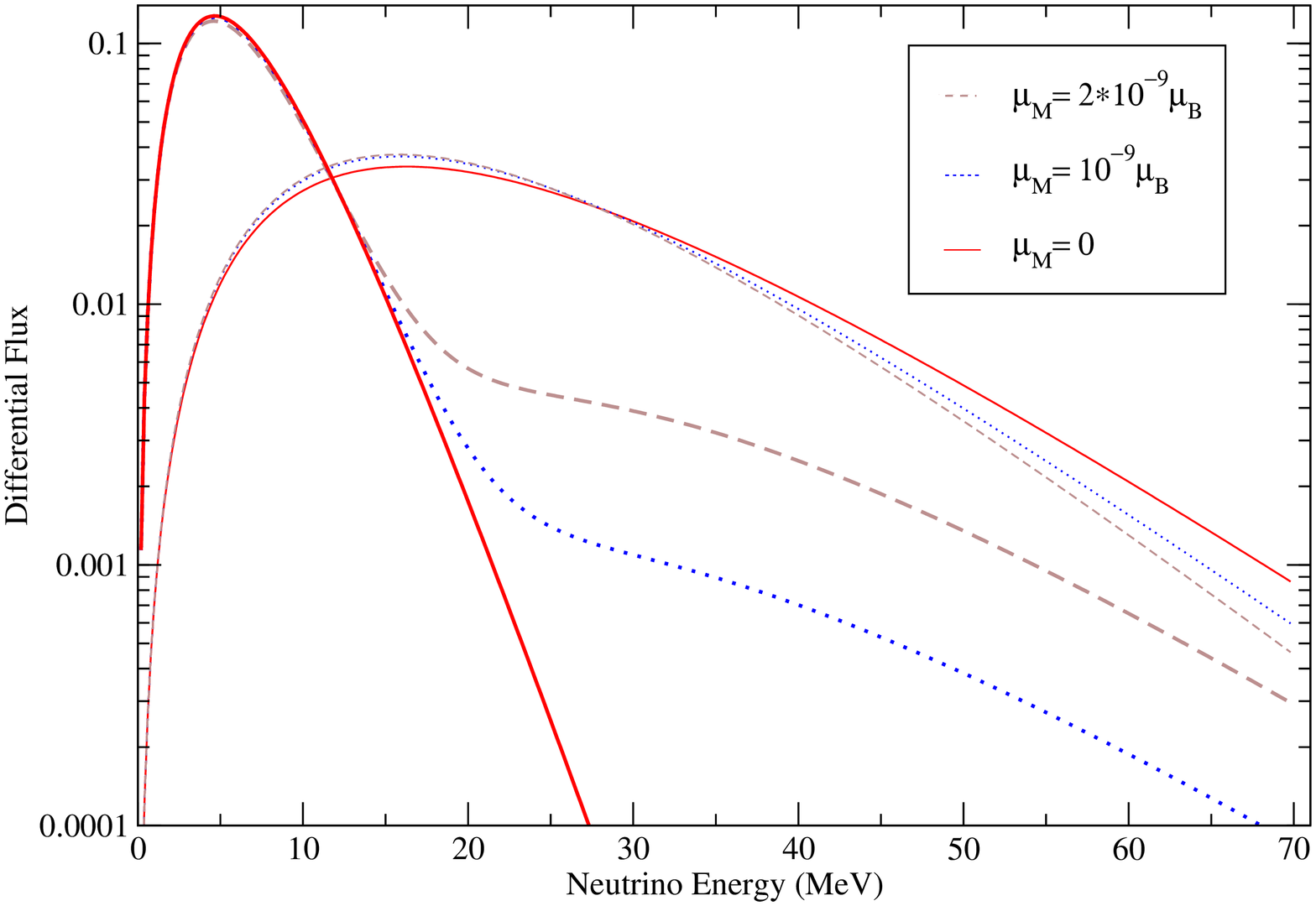}}
\caption{Electron neutrino (thick) and anti-neutrino fluxes (thin) as a function of neutrino energy for the Majorana case. The curves correspond to $\mu_M=0$ (full), $1 \times 10^{-9}$ (dotted) and $2 \times 10^{-9} \mu_B$ (dashed). The case $\mu_M=10^{-10} \mu_B$ is indistinguishable from $\mu_M=0$.}
\label{figfluxes}
\end{figure}

Figure~\ref{figfluxes} 
shows the effect on the electron neutrino and anti-neutrino fluxes
for different values of the neutrino transition magnetic moment
$\mu_M$, defined such that
$\mu_{e\mu}=\mu_{e\tau}=\mu_M$. The transition magnetic moment $\mu_{\mu\tau}$ 
has been fixed to its experimental upper-limit~\cite{Yao:2006px}, 
$2\mu_{\mu\tau}=6.8\times 10^{-10}\mu_B$, as its influence 
on $Y_e(r)$ is very small compared to $\mu_{e\mu}$ and $\mu_{e\tau}$. 
The initial neutrino temperatures correspond to $Y_e(0)=0.30$
({\it cf.} Table~\ref{tableTnu}). From Figure \ref{figfluxes}
one can see that the high energy electron (anti-) neutrino flux tail is enhanced (reduced)
by the active-active conversion with increasing $\mu_M$ while less (more) neutrinos are peaked at low energy.

\begin{figure}
\centerline{\includegraphics[height=8.5cm,angle=-90]{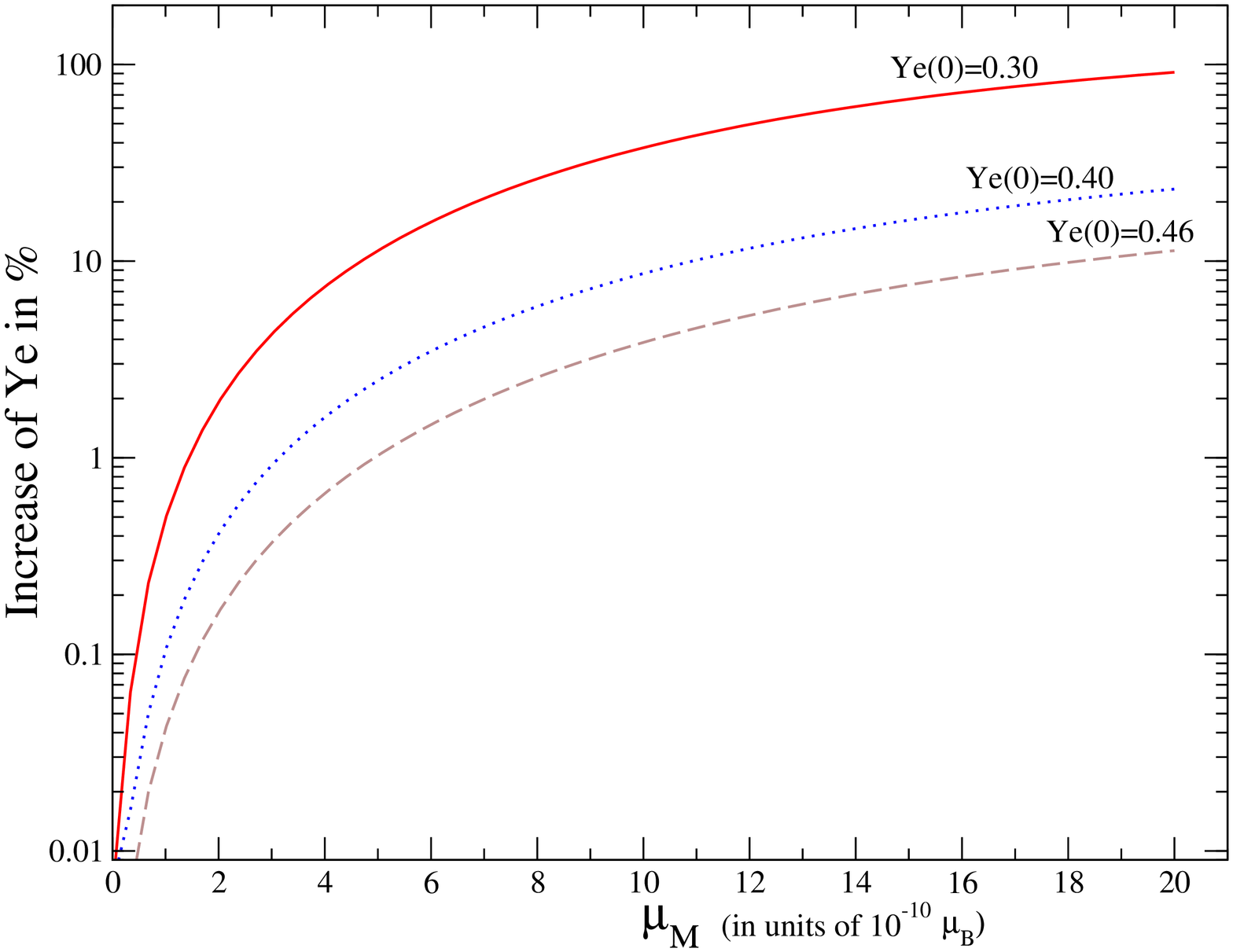}\includegraphics[height=8.5cm,angle=-90]{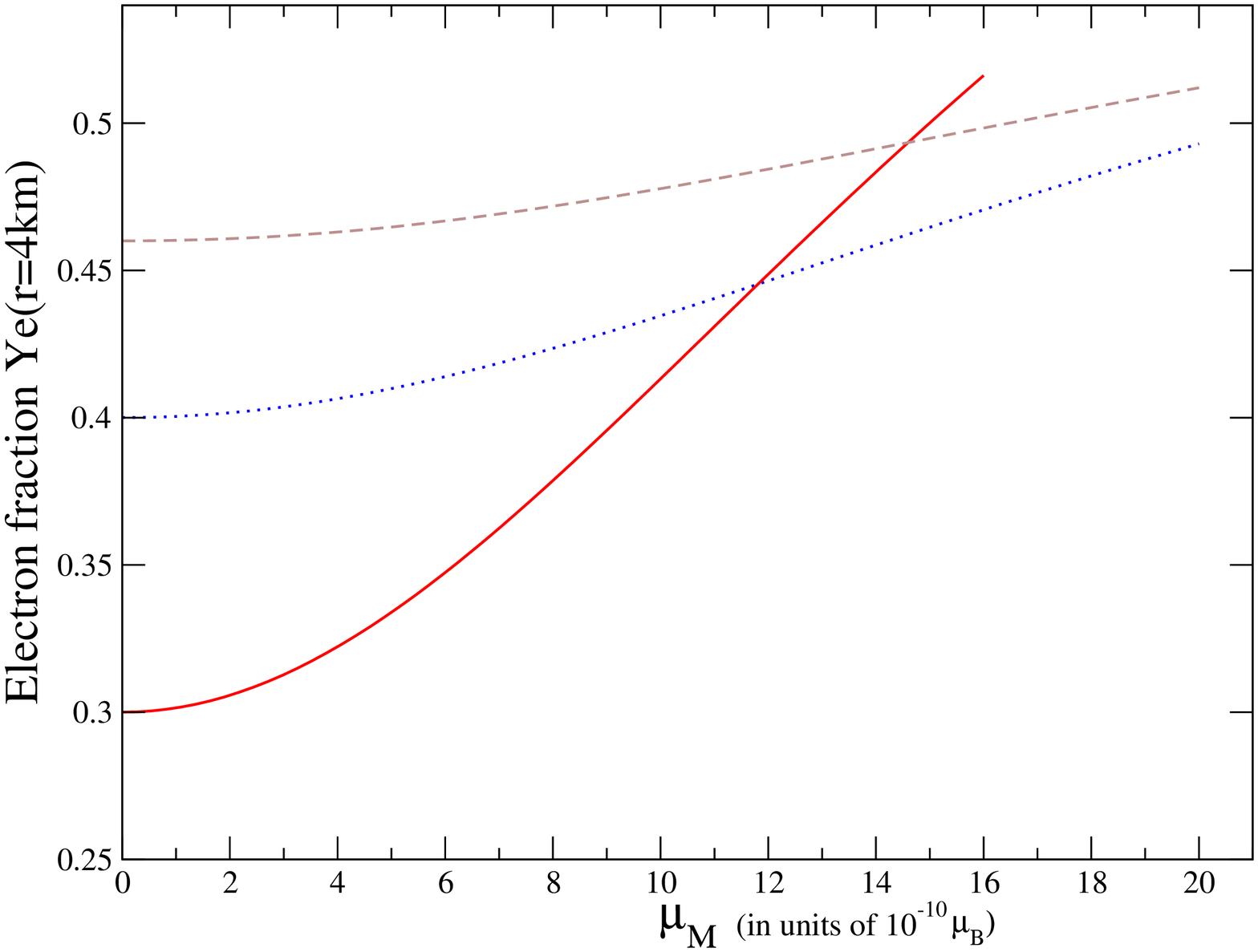}}
\caption{{\sc Case of Majorana neutrinos :} (Left) Increase of the electron
fraction in percentage, $(Y_e(r)-Y_e(r=0))/Y_e(r=0)$, evaluated at a
distance of $r=4$ km from the neutron star surface, as a function of
the magnetic moment $\mu_M=\mu_{e\mu}=\mu_{e\tau}$ (see text). 
(Right) $Y_e(r=4{\rm km})$ as a function of $\mu_M$.}
\label{figYeMajorana}
\end{figure}

Fig~\ref{figYeMajorana} shows the results for the electron fraction
obtained by solving Eq.(\ref{e:15}), as a function of $\mu_M$. 
One can see on both graphs that $Y_e$ increases as $\mu_M$ gets larger 
and that the 
Majorana case shows larger effect on $Y_e$ than the Dirac
case. For transition magnetic moments $\mu_{e\mu}$ and $\mu_{e\tau}$ 
between $1.5\times 10^{-9}\mu_B$ and $2\times 10^{-9}\mu_B$, the electron 
fraction meets the critical value of 0.5 for all $Y_e(0)$ 
(and then all neutrino energy hierarchies).  
The effect on $Y_e$ depends again strongly on the
value of magnetic moments and for $2\mu_M=0.74\times10^{-10}\mu_B$ ({\it i.e.} for
experimental upper-limits), $Y_e$ increases by less then $0.5\%$.

\section{Conclusions}

We pointed out that a
non-zero magnetic moment 
suppresses both the electron neutrino and 
antineutrino fluxes for Dirac neutrinos and 
slightly increases the
electron fraction in a core-collapse Supernova. 
In the Majorana case, 
the high (low) energy neutrino flux tail is enhanced (suppressed)
for a large neutrino 
magnetic moment. Very large values of neutrino magnetic moment also 
increase the initial electron 
fraction. 
However such modifications of the fluxes cannot help reheating the shock 
wave since magnetic moments larger (but uncomfortably close) than the experimental limits are required
to have sizeable effects. 

\section*{ACKNOWLEDGMENTS}
C. Volpe and J. Welzel acknowledge support from the project
ANR-05-JCJC-0023 "Non standard neutrino properties and their impact in
astrophysics and cosmology" (NeuPAC).  The authors acknowledge the
CNRS-Etats Unis 2006 and 2007 grants which have been used during the
completion of this work.  This work was also supported in part by the
U.S. National Science Foundation Grant No.\ PHY-0555231 and in part by
the University of Wisconsin Research Committee with funds granted by
the Wisconsin Alumni Research Foundation.



\end{document}